\documentclass[conference]{IEEEtran}

\usepackage{graphicx}  
\begin{document}

\title{Phase Noise Modeling of Opto-Mechanical Oscillators}

\author{\authorblockN{Siddharth Tallur, Suresh Sridaran, Sunil A. Bhave}
\authorblockA{OxideMEMS Lab, School of Electrical and Computer Engineering\\
Cornell University\\
Ithaca, New York 14853\\
Email: sgt28@cornell.edu}
\and
\authorblockN{Tal Carmon}
\authorblockA{University of Michigan\\
Ann Arbor, Michigan 48109}
}

%


\maketitle

\begin{abstract}
We build upon and derive a precise far from carrier phase noise model for radiation pressure driven opto-mechanical oscillators and show that calculations based on our model accurately match published phase noise data for such oscillators. Furthermore, we derive insights based on the equations presented and calculate phase noise for an array of coupled disk resonators, showing that it is possible to achieve phase noise as low as -80 dBc/Hz at 1 kHz offset for a 54 MHz opto-mechanical oscillator.
\end{abstract}


%
\IEEEpeerreviewmaketitle

\section{Introduction}
Opto-mechanical oscillators (OMOs) are unique as they feature optical input and output for both power and signal. Unlike traditional temperature compensated crystal oscillators (TCXOs), their performance is not limited by the $f \times Q$ product of the material and they are expected to demonstrate narrower linewidth and better phase noise. Based on origin of self-sustaining oscillations, OMOs can be classified into 3 main categories: radiation pressure\cite{refequation2}\cite{ref1}\cite{ref1.1}, gradient force\cite{ref2} and stimulated Brillouin scattering\cite{ref3}, as shown in Figure \ref{fig_omos} below.

\begin{figure}[htbp]
\centering
\includegraphics[width = 3in]{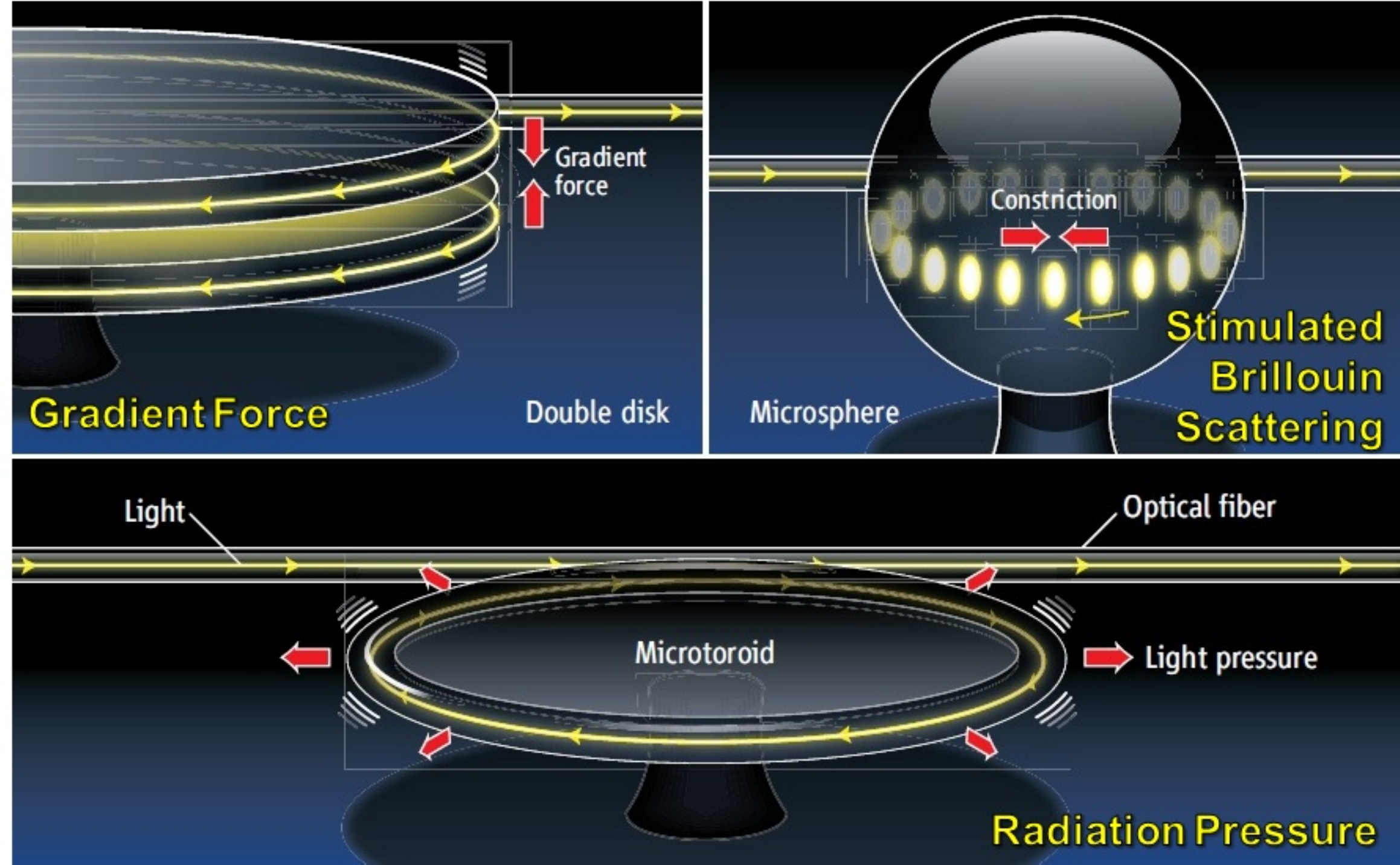}
\caption{Illustration of various kinds of opto-mechanical oscillators \cite{refomos}}
\label{fig_omos}
\end{figure}

\hspace*{0.2 cm}While experimental results have confirmed narrow linewidths for OMOs, experimental and theoretical studies of phase noise in literature are still limited. In this work, we build upon the existing theoretical models and derive a more precise phase noise model for radiation pressure driven OMOs demonstrated in \cite{ref1}. We start with an overview of the radiation pressure driven opto-mechanical oscillator. Then we present analysis of the linewidth and phase noise in these oscillators. In the subsequent sections we propose and analyse an array of mechanically coupled disks that is predicted to have better phase noise based on insights derived from our model.

\section{Radiation pressure driven opto-mechanical oscillators}

\begin{figure}[htbp]
\centering
\includegraphics[width = 3.2in]{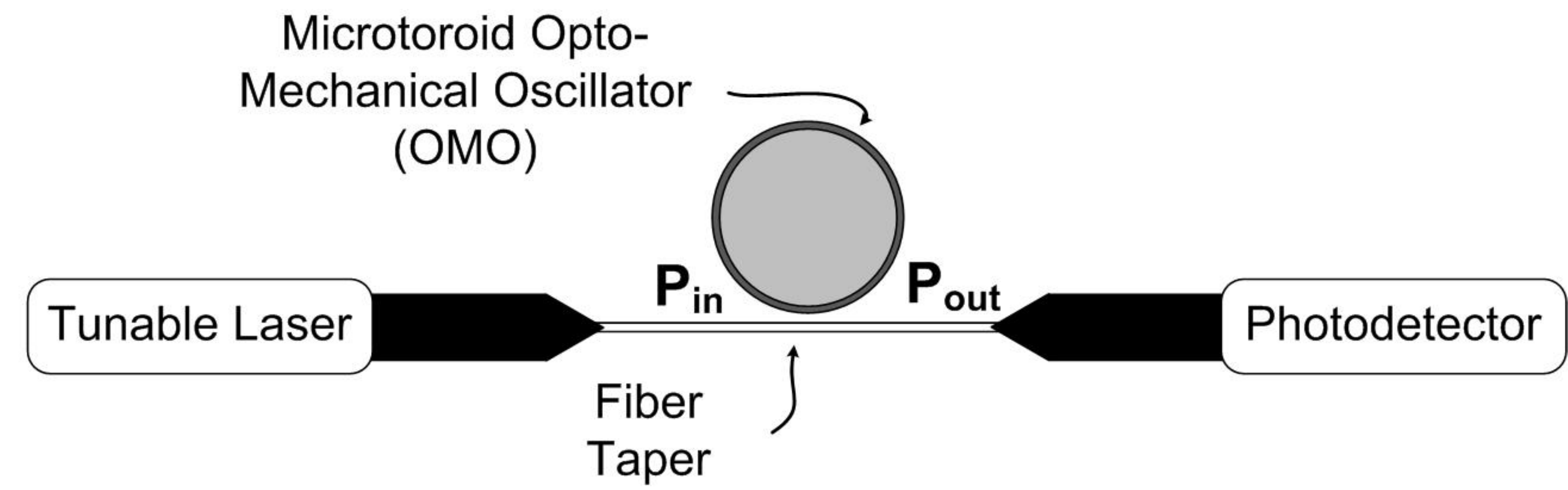}
\caption{Illustration of setup for radiation pressure driven OMO}
\label{fig_setup}
\end{figure}

Figure \ref{fig_setup} shows a schematic of the device presented in \cite{ref1}.

The structure consists of a microtoroid resonator in close proximity with a fiber tapered waveguide. Light is fed into the waveguide from the tunable laser and the photodetector reads off the output.  Light couples into the resonator from the waveguide through evanescent coupling. The geometry of the disk is chosen such that the optical path inside the resonator is an integer multiple of the wavelength of the light. Due to the high quality factor, the intensity of light builds up to very large values enabling strong optomechanical interactions. The resulting radiation pressure of the circulating optical power exerts a radial force on the microtoroid that causes the structure to expand. However, this causes a change in the radius and hence the optical path inside the cavity which causes a shift in the resonant frequency. This leads to decrease in circulating optical power thereby reducing the radiation pressure. The restoration of the mechanical deformation sets up oscillations in the radial displacement. Usually evanescent coupling through a fiber taper is used to couple light into and out of the resonator.
%

\section{Linewidth and Phase noise in OMOs}
The classical Leeson model for electronic oscillators\cite{refleeson} relates the phase noise of an oscillator to its linewidth. We extend this well established theory to the case of opto-mechanical oscillators.

\subsection{Oscillation linewidth}
Mani et al. have characterized the mechanical oscillation linewidth as follows\cite{ref1}:

\begin{equation}\label{eqnmani}
 \Delta \Omega = \frac{1}{2\pi} \left(\frac{4k_BTQ_{tot}^2}{m_{eff}\Omega_0^2R_0^2}\right)\left(\frac{\Gamma^2\Delta\Omega_0}{M^2}\right)
\end{equation}

The linewidth is set by $k_B$ (Boltzmann constant), $T$ (absolute temperature), $Q_{tot}$ (total optical Q), $m_{eff}$ (effective mass of mechanical resonance), $\Omega_0$ (mechanical resonant frequency), $R_0$ (disc radius), $\Gamma$ (optical modulation transfer function\cite{ref1}), $\Delta\Omega_0$ (intrinsic mechanical oscillation linewidth) and $M$ (modulation depth). $\Gamma$ represents the finite response time taken by the resonator to respond to modulation of the optical path length. Intuitively, it should seem that $\Gamma$ is a low pass filter function of oscillation frequency, i.e. at higher mechanical oscillation frequencies its value is smaller. $\Gamma$ and $M$ are experimental parameters. However, for reasons we discuss later, we will write the expression for the linewidth in terms of more fundamental parameters describing the system:

\begin{equation}\label{eqnlw}
 \Delta \Omega = \frac{1}{2\pi} \left(\frac{4k_BT}{m_{eff}\Omega_0^2}\right)\left(\frac{\Delta\Omega_0}{r^2}\right)
\end{equation}

where $r$ is the amplitude of radial displacement.
\subsection{Phase noise}

%
%

\begin{figure}[htbp]
\centering
\includegraphics[width = 2.6in]{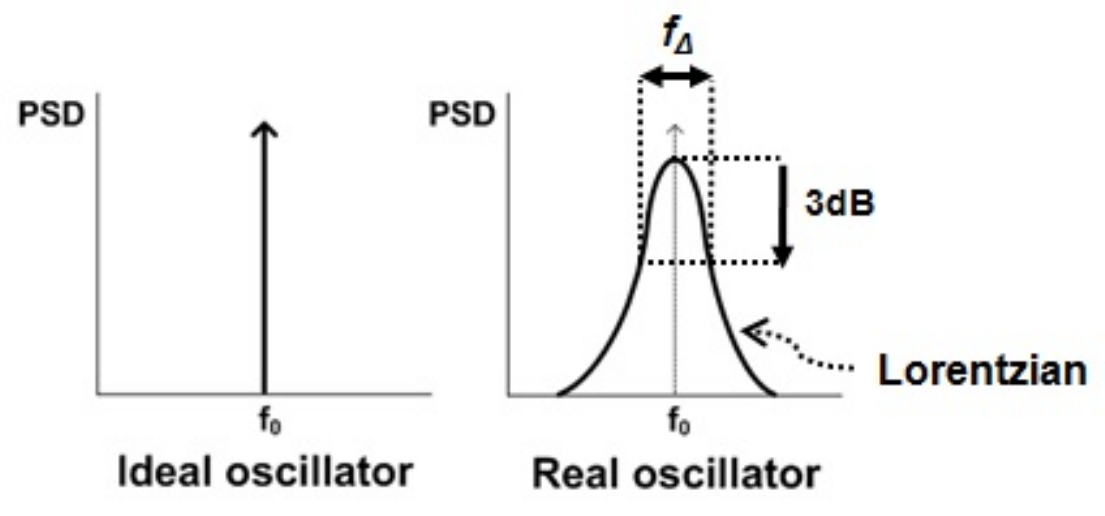}
\caption{Illustration of spectral width widening due to noise in oscillators}
\label{fig_psd}
\end{figure}

Phase noise is frequency domain representation of random fluctuations in the phase of a waveform. An ideal oscillator generates a pure sine wave that can be represented as a pair of delta functions at the oscillation frequency in the frequency domain, i.e. all the signal power is at a single frequency. However, due to presence of various noise sources within the oscillation loop, real oscillators demonstrate time domain instabilities , which appear as random fluctuations in the spacing between zero crossings of the waveform from the half period of oscillation. This can be interpreted as random fluctuations in the frequency of the oscillator, as illustrated in Figure \ref{fig_psd}.

\begin{figure}[htbp]
\centering
\includegraphics[width = 2.6in]{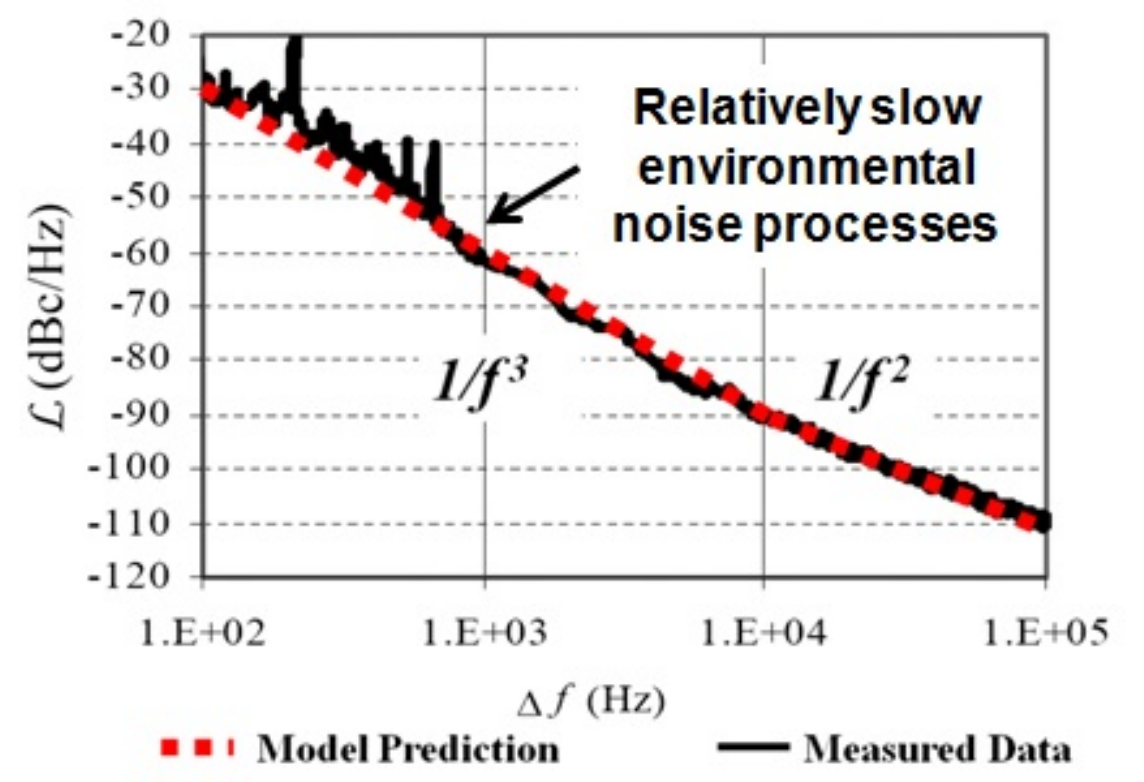}
\caption{Comparison of model prediction to measured phase noise data \cite{ref1}}
\label{fig_modelmeas}
\end{figure}

Thus, in the power spectral density (PSD) of the oscillator, the oscillation energy is not located at a single frequency but spreads to adjacent frequencies which can be characterized by the linewidth of the oscillator. The shape of this spectrum for typical oscillators is a Lorentzian distribution, which is described by the following equation

\begin{equation}
 S_{v}(f_{0}+\Delta f) = \frac{A^2}{2\pi} \frac{f_{\Delta}}{f_{\Delta}^2 + (\Delta f)^2}
\end{equation}

where $A^2/2$ is the signal power at the oscillation frequency and $f_{\Delta}$ is the linewidth. $\Delta f$ specifies the offset from oscillation frequency $f_{0}$. Phase noise, which is specified in dBc/Hz (decibels below carrier per hertz), is one metric used to quantify the amount of signal power that is contained in close offsets from the desired signal frequency. This way of quantifying phase noise gives a measure of the spread of oscillation power to frequencies offset from the center (carrier) frequency\cite{refpn}.\\
\hspace*{0.2 cm}Phase noise is one of the most important characteristics of a self-sustained oscillator. An example of its importance can be seen in the use of oscillators as frequency references in communication systems: the phase noise of the oscillator impacts the bit error rate and security. Also, oscillators do not have memory of phase of the waveform and as such are unable to restore phase, which leads to accumulation of phase deviations caused by phase noise, or drift. Hence it is crucial to be able to predict phase noise of an oscillator during the design phase.\\
\hspace*{0.2 cm}We can use Leeson's model to model the phase noise of OMOs using the expression of linewidth from equation \ref{eqnlw}. The following equation gives an expression for phase noise in both $1/f^2$ and $1/f^3$ regimes. $\omega_{1/f^3}$ is the corner frequency for $1/f^3$ noise. The $1/f^3$ noise is attributable to slow environmental noise processes \cite{ref1}.
\begin{eqnarray}
 L(\Delta f) &=& 10log_{10}\left(\frac{1}{2\pi}.\frac{\Delta \Omega}{\Delta f^2}\right) \ldots 1/f^2     \nonumber \\
   &=& 10log_{10}\left(\frac{1}{2\pi}.\frac{\Delta \Omega}{\Delta f^2}.\frac{\omega_{1/f^3}}{\Delta f}\right) \ldots 1/f^3    
\end{eqnarray}
\hspace*{0.2 cm}Figure \ref{fig_modelmeas} shows that this model gives a close match to the measured data.

%
Based on this model, we can draw some insights into designing oscillators with low phase noise. As can be seen from the equations for linewidth and phase noise, we see that if there could be a way to increase the effective mass of the system, $m_{eff}$, while ensuring that the amplitude of radial displacement, $r$, does not drop so much that $\frac{1}{m_{eff}r^2}$  increases, then we can improve the phase noise of the system. However, the system becomes stiffer and hence we may need larger input power to obtain the same radial displacement. We propose a solution to this problem by using multiple disks mechanically coupled to each other. Such a configuration allows us to actuate multiple disks, thereby maintaining operating power.

\section{Mechanically coupled opto-mechanical oscillators}

\subsection{Dynamics of radiation pressure driven OMO}
\begin{figure}[htbp]
\centering
\includegraphics[width = 2.6in]{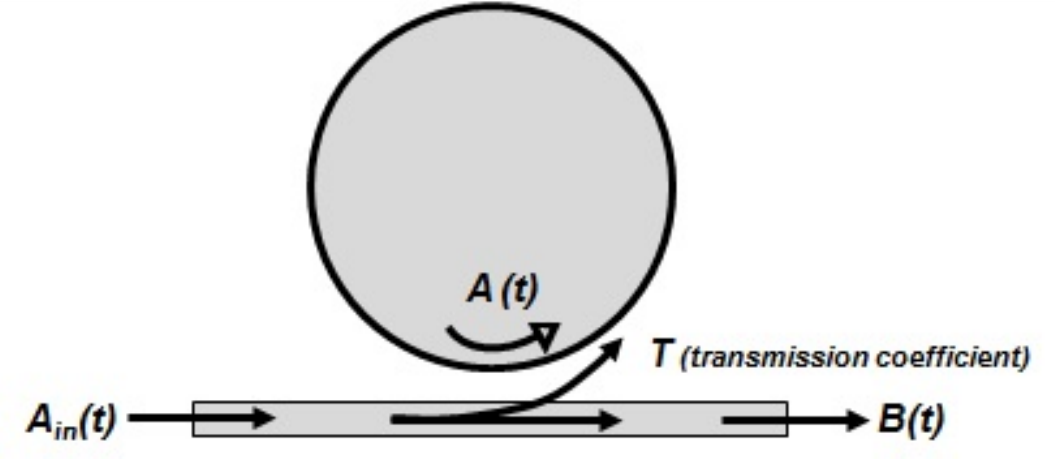}
\caption{Illustration of opto-mechanical power coupling}
\label{fig_powerdiag}
\end{figure}

The differential equations governing the dynamics for a radiation pressure driven OMO have been discussed in previous publications \cite{refequation2}\cite{refequation}.

\begin{equation}
\ddot{r}(t) + \gamma_0\dot{r}(t) + \Omega^2r(t) = \frac{2\pi n}{m_{eff} c}|{A(t)}|^2
\end{equation}

\begin{equation}
 \dot{A}(t) + A(t)\left[\frac{\omega}{2Q_{tot}} + i\Delta \omega(t)\right] = iB(t)\sqrt[]{\frac{\omega}{TQ_{ext}}}
\end{equation}

where $r$ is the radial displacement amplitude and $A_{in}$ and $B$ are input and output optical power intensities as shown Figure \ref{fig_powerdiag}. $A$ is power inside the resonator.

The output power $B(t)$ for a disk can be specified in terms of its input power $A_{in}(t)$ and the power inside the resonator $A(t)$ as follows

\begin{equation}
 B(t) = \left(1-\frac{\tau_{0}\omega}{2Q_{tot}}\right)A_{in}(t) + i\sqrt[]{\frac{\tau_{0}\omega}{Q_{tot}}}A(t)
\end{equation}

where $\tau_{0}$ is the time taken by light to complete one circulation of the resonator. When the laser is detuned by $\Delta\omega_0$ relative to the optical mode resonance, the optical detuning $\Delta\omega(t)$ is given by the following relation

\begin{equation}
 \Delta \omega(t) = \Delta \omega_0 + \left(\frac{\omega}{R_0}\right)r(t)
\end{equation}

%

\subsection{Mechanically coupled OMOs}
\begin{figure}[htbp]
\centering
\includegraphics[width = 3in]{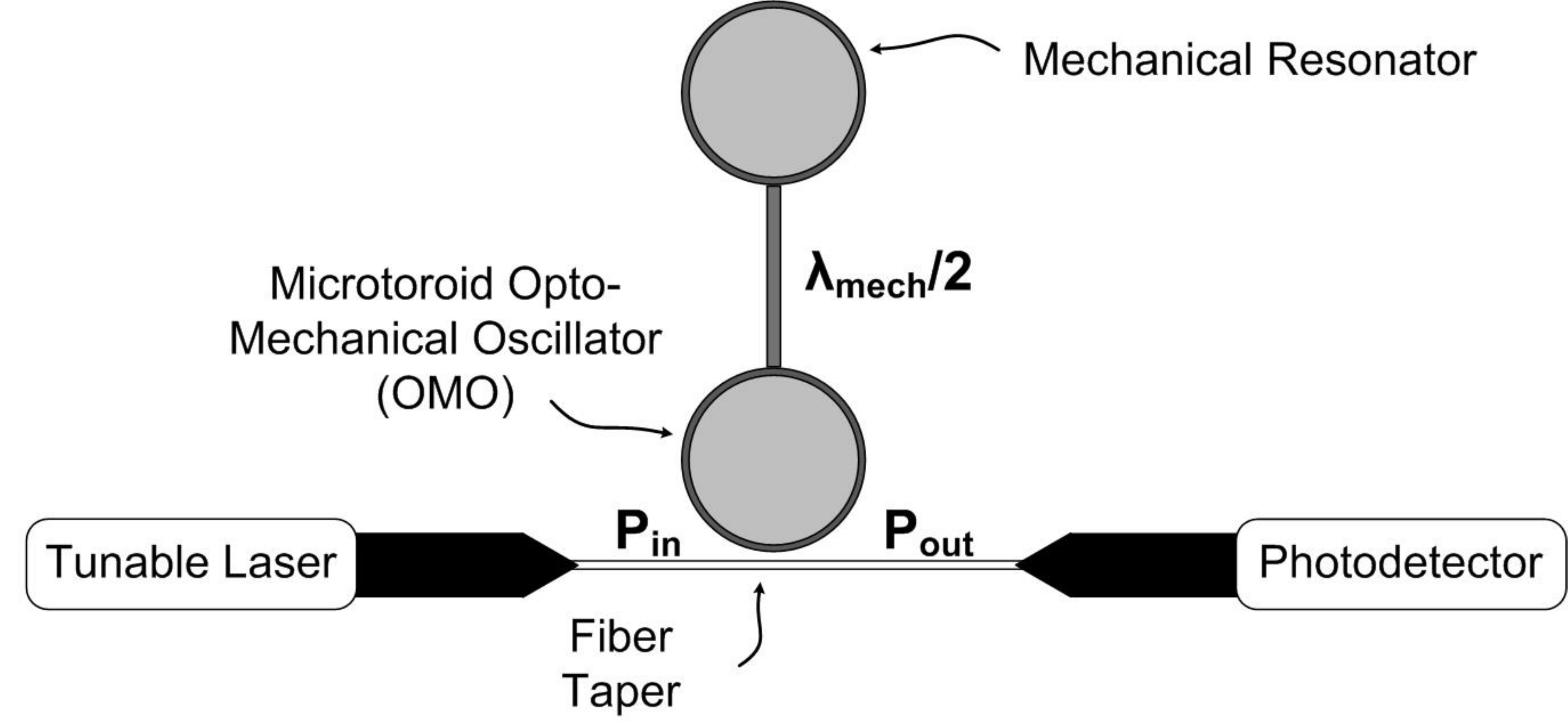}
\caption{Proposed array of disks}
\label{fig_proponetowg}
\end{figure}

We propose a system of disks mechanically coupled to each other as shown in Figure \ref{fig_proponetowg}, where only one resonator is coupled to the waveguide. The figure only shows two coupled disks, but one can visualize a linear array of disks coupled in a similar fashion.

The coupling beam length is equal to $\lambda_{mech}/2$, where $\lambda_{mech}$ corresponds to the wavelength of the mechanical oscillation frequency. This beam ensures strong coupling between oscillations of both disks. This configuration increases the effective mass for the system while maintaining the oscillation frequency. However due to increased stiffness of the system, a higher laser power is necessary to maintain the radial displacement amplitude. 

%

\begin{figure}[htbp]
\centering
\includegraphics[width = 3.6in]{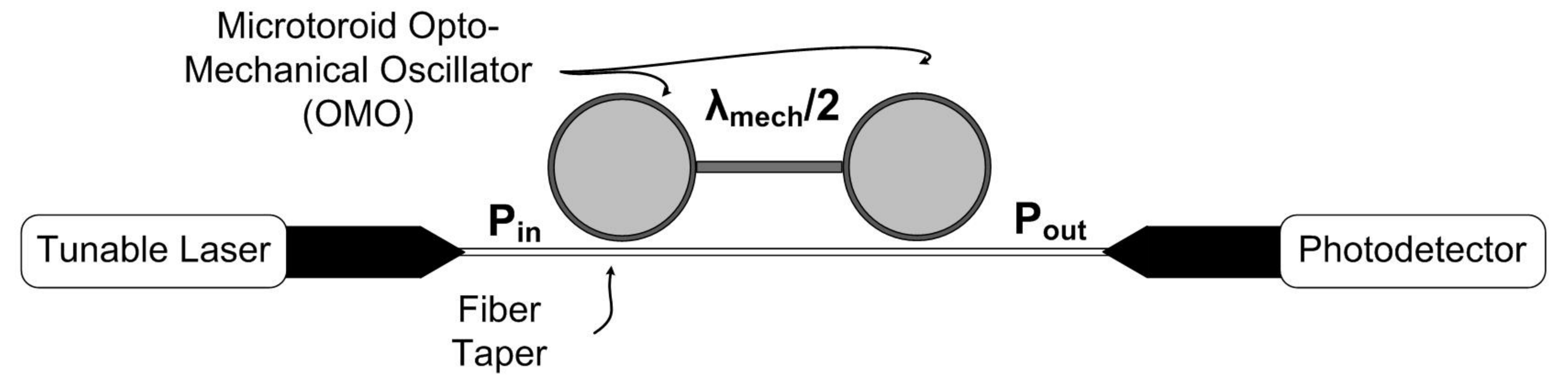}
\caption{Proposed array of disks with multiple disks actuated}
\label{fig_propalltowg}
\end{figure}

\hspace*{0.2 cm}To overcome this high power requirement, we can actuate multiple disks instead of a single one. Figure \ref{fig_propalltowg} illustrates how this can be achieved.

\subsection{Dynamics of mechanically coupled OMOs}
\begin{figure}[htbp]
\centering
\includegraphics[width = 3in]{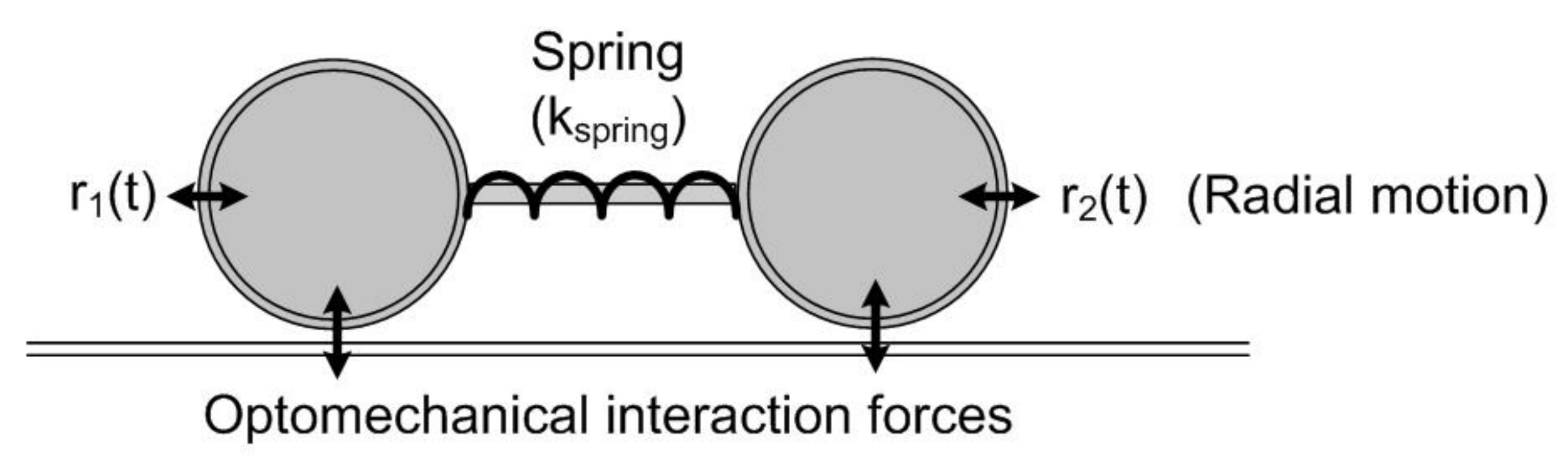}
\caption{Illustration of model for two coupled disk system}
\label{fig_model}
\end{figure}

The model for a single disk OMO \cite{refequation2} can be extended to the mechanically coupled OMO system. The model for a two disk system can be interpreted as described in Figure \ref{fig_model}. The stiffness of the coupling beam is modeled by an equivalent spring of spring constant $k_{spring}$.

The dynamics of this system can be described by the following set of equations:

\begin{eqnarray}\nonumber
\ddot{r}_1(t) + \gamma_0\dot{r}_1(t) + \Omega^2r_1(t)&=& \frac{k_{spring}}{m_{eff}}[-r_1(t)+r_2(t)]\\
&& + \frac{2\pi n}{m_{eff} c}|{A_1(t)}|^2
\end{eqnarray}

\begin{eqnarray}\nonumber
\ddot{r}_2(t) + \gamma_0\dot{r}_2(t) + \Omega^2r_2(t)&=& \frac{k_{spring}}{m_{eff}}[r_1(t)-r_2(t)]\\
&& + \frac{2\pi n}{m_{eff} c}|{A_2(t)}|^2
\end{eqnarray}

\begin{equation}
 \dot{A}_1(t) + A_1(t)\left[\frac{\omega}{2Q_{tot}} + i\Delta \omega_1(t)\right] = iB_1(t)\sqrt[]{\frac{\omega}{TQ_{ext}}}
\end{equation}

\begin{equation}
 \dot{A}_2(t) + A_2(t)\left[\frac{\omega}{2Q_{tot}} + i\Delta \omega_2(t)\right] = iB_2(t)\sqrt[]{\frac{\omega}{TQ_{ext}}}
\end{equation}

The optical detuning for the two disks can be described as follows

\begin{equation}
 \Delta \omega_1(t) = \Delta \omega_0 + \left(\frac{\omega}{R_0}\right)r_1(t)
\end{equation}

\begin{equation}
 \Delta \omega_2(t) = \Delta \omega_0 + \left(\frac{\omega}{R_0}\right)r_2(t)
\end{equation}

\begin{figure}[htbp]
\centering
\includegraphics[width = 3in]{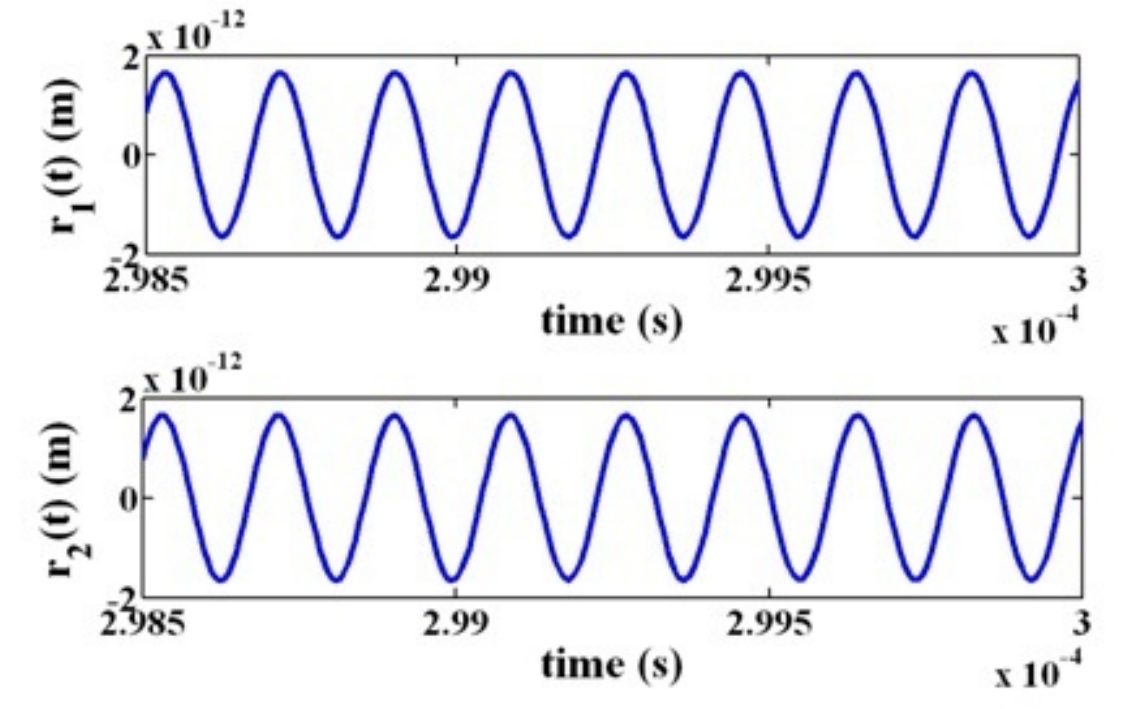}
\caption{Radial displacement of both disks - one disk coupled to waveguide: amplitude of $r_{1} = r_{2} =$ 1.65pm}
\label{fig_numsim_one}
\end{figure}

\begin{figure}[htbp]
\centering
\includegraphics[width = 3in]{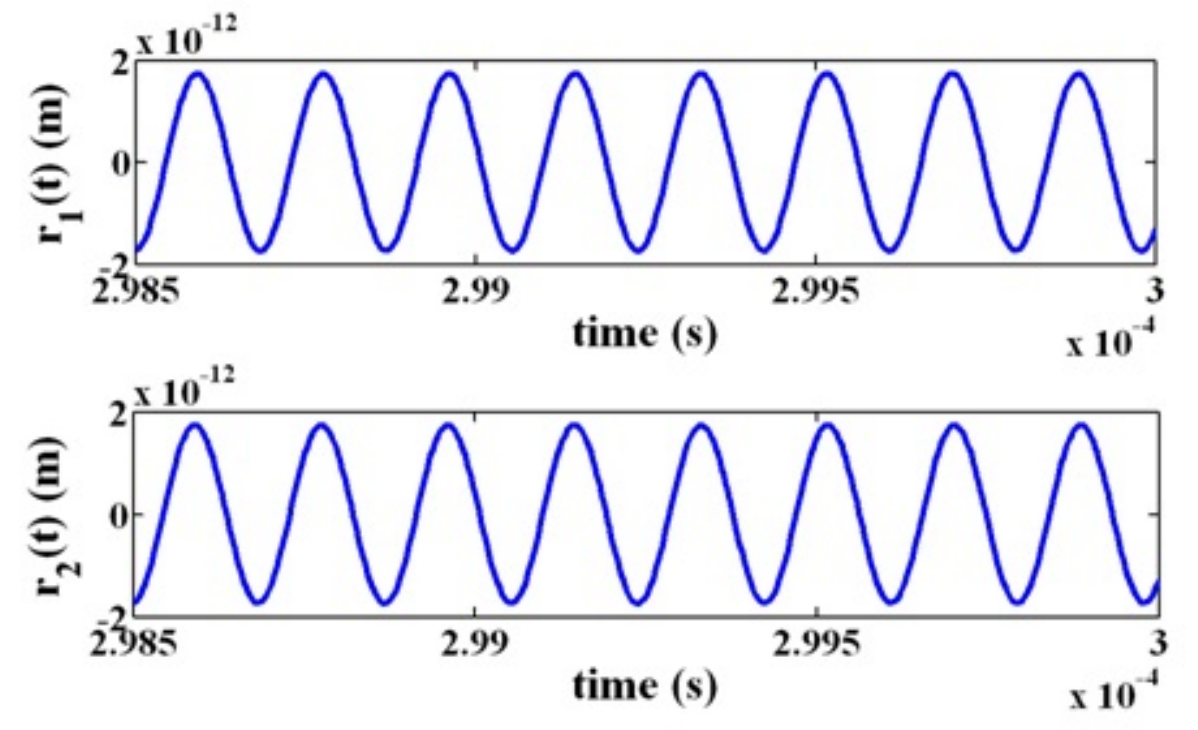}
\caption{Radial displacement of both disks - both disks coupled to waveguide: amplitude of $r_{1} = r_{2} =$ 1.93pm}
\label{fig_numsim}
\end{figure}

This system was numerically simulated in MATLAB. The half wavelength coupling beam presents infinite stiffness in the system \cite{sslithesis}. However, since it is difficult to model infinite stiffness in time domain numerical simulations, we have used a large value for $k_{spring}$ in our simulations. Figure \ref{fig_numsim_one} shows numerical simulation results for a 2 disk system with only one disk coupled to the waveguide. Figure \ref{fig_numsim} shows numerical simulation results for a 2 disk system with both disks coupled to the waveguide. As can be seen, the amplitude of radial oscillations is higher for the case where both disks are coupled to the waveguide.

%
%

This model can be extended to more than two coupled resonators. Table \ref{tabmr2} shows radial displacement waveforms for a system of five coupled disks after settling.

\begin{table}[htbp]
\renewcommand{\arraystretch}{1.3}
\caption{Comparison of stored energy metric $m_{eff}r^2 (10^{-34} kg-m^2)$ for different number of disks}
\label{tabmr2}
\begin{center}
\begin{tabular}{|c|c|}
\hline
Number of disks & Stored energy metric\\
\hline
1 & 1.56\\
\hline
2 & 3.72\\
\hline
3 & 4.08\\
\hline
5 & 6.12\\
\hline
\end{tabular}
\end{center}
\end{table}


\section{Phase noise for mechanically coupled OMO}
Having seen that a system of coupled disks can oscillate in synchrony, we can use the phase noise model to estimate the phase noise for this system. We can extend the design presented in \cite{ref1} to an array of 100 such disks mechanically coupled to each other. Using the model we obtain the following data for phase noise at different carrier offset frequencies, as shown in the table, showing an overall $20 dBc/Hz$ improvement due to increased effective mass.

\begin{table}[htbp]
\renewcommand{\arraystretch}{1.3}
\caption{OMO Phase noise (dBc/Hz)}
\label{table_example}
\begin{center}
\begin{tabular}{|c|c|c|}
\hline
Frequency offset (Hz) $\downarrow$ & Single disk\cite{ref1} & Proposed design\\
\hline
100 & -30 & -50\\
\hline
1000 & -60 & -80\\
\hline
10000 & -90 & -110\\
\hline
100000 & -110 & -130\\
\hline
\end{tabular}
\end{center}
\end{table}

\section{Conclusion}
We have presented a model for predicting phase noise for radiation pressure driven OMOs. Insights derived from this model will enable one to design low phase noise oscillators. Furthermore, we have proposed and analysed an array of disks mechanically coupled to each other and have demonstrated better phase noise for the design using the model. The phase noise for this array is modeled to be $-80 dBc/Hz$ at $1 kHz$ offset for a $54 MHz$ opto-mechanical oscillator. This suggests that insight gained from the proposed phase noise model can help us achieve improved phase noise performance for chip scale opto-mechanical devices. More effort has to be spent on understanding the frequency scalability of this model, namely understanding how the parameters  $\Gamma$ and $M$ in equation \ref{eqnmani} vary with frequency.


\section*{Acknowledgment}
The authors would like to thank Prof. Ehsan Afshari for interesting discussions on phase noise in oscillators. We would also like to express our thanks to Eugene Hwang for providing insights into modeling of the coupled OMO system.




\begin{thebibliography}{1}
  
  \bibitem {refequation2}
Carmon T., Rokhsari H., Yang L., Kippenberg T. J., Vahala K. J., \emph{Temporal behavior of radiation-pressure-induced vibrations of an optical microcavity phonon mode}\hskip 1em plus
  0.5em minus 0.4em\relax Phys. Rev. Lett. 94, 223902, 2005.
  
\bibitem {ref1}
Hossein-Zadeh M., Rokhsari H., Hajimiri A., Vahala K. J. \emph{Characterization of a radiation-pressure-driven micromechanical oscillator}\hskip 1em plus
  0.5em minus 0.4em\relax Phys. Rev. A 74, 023813, 2006
  
  \bibitem {ref1.1}
Kippenberg T. J., Vahala K. J. \emph{Cavity opto-mechanics}\hskip 1em plus
  0.5em minus 0.4em\relax Opt. Express 15, 17172–17205, 2007.
  
\bibitem {ref2}
Eichenfeld M., Michael C. P., Perahia R., Painter O. \emph{Actuation of micro-optomechanical systems via cavity enhanced
optical dipole forces}, 3rd~ed.\hskip 1em plus
  0.5em minus 0.4em\relax Nature Photonics 1, 416–422, 2007.
  
\bibitem {ref3}
Tomes M., Carmon T., \emph{Photonic micro-electromechanical systems vibrating at X-band (11-GHz) rates}, 3rd~ed.\hskip 1em plus
  0.5em minus 0.4em\relax Phys. Rev. Lett. 102, 113601, 2009.

\bibitem {refomos}
Cho A., \emph{Putting Light's Light Touch to Work As Optics Meets Mechanics}\hskip 1em plus
  0.5em minus 0.4em\relax Science Vol. 328. no. 5980, 812 - 813, 2010.
  
\bibitem {refpn}
Hajimiri A., Lee T., \emph{Oscillator Phase Noise: A Tutorial}\hskip 1em plus
  0.5em minus 0.4em\relax IEEE Journal of Solid-State Circuits, vol. 35, pp. 326–336, Mar. 2000.
  
\bibitem {refleeson}
Leeson D. B., \emph{A simple model of feedback oscillator noise spectrum}\hskip 1em plus
  0.5em minus 0.4em\relax Proc. IEEE, vol. 54, pp. 329–330, Feb. 1966,

\bibitem {refequation}
Rokhsari H., Kippenberg T., Carmon T., Vahala K. J., \emph{Theoretical and experimental study of
radiation pressure-induced mechanical oscillations (parametric instability) in optical microcavities}\hskip 1em plus
  0.5em minus 0.4em\relax IEEE J. Sel. Top. Quantum Electron. 12, 96–107, 2006.
  
\bibitem {sslithesis}
Li S.-S., Lin Y.-W., Ren Z., Nguyen C. T.-C. , \emph{An MSI micromechanical differential disk-array filter}, Transducers 2007,\hskip 1em plus
  0.5em minus 0.4em\relax pp. 307-311.
  
  
\end{thebibliography}
%

\end{document}